# OY Car During Normal Outburst: Balmer Emission From The Red Star And The Gas Stream

E. T. Harlaftis[1] and T. R. Marsh[2]

[1] Isaac Newton Group of Telescopes, Apartado de Correos 321, Santa Cruz de La Palma, E-38780 Canary Islands, Spain.
[2] Department of Physics, University of Southampton, Southampton, SO17 1BJ, UK.

**Abstract:**

We present observations of OY Car, obtained with the Anglo-Australian Telescope, during a normal outburst in August 1991. Two sinusoidal components are resolved in the H$\beta$ trailed spectra and we determine the location of the narrow component to be on the secondary star with a maximum contributed flux of $\sim$2.5 per cent to the total flux. Imaging of the line distribution reveals that the other emission component is associated with the gas stream. This follows a velocity close to the ballistic one from the red star to a distance of $\sim$0.5 R$_{L_1}$ from the white dwarf. This emission penetrates the accretion disc (from 0.5–0.1 R$_{L_1}$), with a velocity now closer (but lower) to the keplerian velocities along the path of the gas stream. We finally discuss the implications of having observed simultaneously line emission from the gas stream and the red star during outburst.

**Keywords:** stars: binaries: eclipsing – stars: cataclysmic variables – stars: individual: OY Car

## 1. Introduction

The secondary stars are often undetectable in cataclysmic variables and this is particularly true for systems of short orbital period in which the secondary stars are especially dim (M dwarfs), but also for high mass transfer systems where light from the accretion disc dominates (nova-like variables). In recent years, line emission originating from the red star has been detected in a few cataclysmic variables (UX UMa, Schlegel et al. 1983; RW Tri, Kaitchuck et al. 1983; SS Cyg, Hessman 1986; IX Vel, Beuermann & Thomas 1990; RX And, Kaitchuck, Mansperger and Hantzios 1988; IP Peg, Marsh and Horne 1990; U Gem, Marsh et al. 1990; DW UMa, Dhillon et al. 1994; YZ Cnc, Harlaftis et al. 1995) providing a new approach in studying the secondary star. Indeed, there are cases where the secondary star can be detected and subsequently studied only by its line emission (e.g. IX Vel; Beuermann and Thomas 1990).

It was realized by us that such narrow emission lines from the secondary star may not have been resolved before and that with adequate resolution these lines could be detected in the trailed spectra of many more systems than previously thought. Indeed, no emission component had been found before from the secondary star of OY Car. Here, we show such a detection which springs from the higher than normal resolution of our observations and the outburst state of the system.

## 2. Observations and Data Reduction

OY Car was observed with the 3.9-m Anglo-Australian Telescope for a full orbital cycle on the night 31 August 1991 while the target was at maximum of a normal outburst (maximum was 12.1 mag on 30 August and lasted for one day; Frank Bateson, private communication). The orbital cycle was covered with 78 spectra of 1 minute long exposure (dead time between exposures was 10 seconds). The RGO spectrograph with the Thomson CCD and a grating of 1200 lines mm$^{-1}$ gave a wavelength coverage from 4585–4965 Å at FWHM of 1.26 Å (or $\sim$80 km s$^{-1}$ at H$\beta$) resolution.

The slit was rotated so that a comparison star was placed in the slit for later slit loss corrections (seeing was 2.3 arcseconds). Wide slit photometric spectra of a flux standard and of the comparison star were also obtained. Comparison arc spectra were taken every 30 min.

After the bias level was removed from the data images and the sky contribution subtracted from the spectra, the object and comparison spectra were optimally extracted (Marsh 1989). For the wavelength calibration of the spectra, interpolation was used between neighbouring arc spectra. The root mean square of the polynomial fits is ~0.014 Å. Although we corrected for slit losses, the flux calibration is only accurate to 25 per cent mainly because some observations were seriously affected by clouds and high air-mass (the latter resulted because the observations were done in an unfavourable month for OY Car). The ephemeris we used for the orbital phase determination is the following:

HJD = 2443993.552733 + 0.0631209239 E

as derived from Space Telescope observations of OY Car and accurate to ~1 second (Marsh et al. 1995).

## 2.1 The Average Spectrum

We display the average out-of-eclipse spectrum of OY Car in Fig. 1. The blue spectrum shows double-peaked H$\beta$ and He I emission profiles with prominent absorption cores going below the continuum level. High excitation lines such as the C III/N III and He II $\lambda$4686 are in emission, apparently forming double-peaked profiles with an absorption core component. This looks similar to spectra of IP Peg during outburst although there the He II 4686 is stronger than the H$\beta$ line (Marsh & Horne 1990). This is possibly related to the higher mass-transfer observed in long period cataclysmic variables which increases the temperature of the photoionizing source. Unfortunately, there are no other optical spectra of OY Car or Z Cha in outburst and very few in superoutburst to compare with. A few H$\alpha$ spectra at the very end of a superoutburst show a stronger red peak to the blue one and an absorption core which is well above the continuum level (Hessman et al. 1992). Spectra of Z Cha in superoutburst show the same characteristics as OY Car in outburst (Vogt 1982).

Comparison with spectra of OY Car during quiescence shows that the main difference is the absence of the broad absorption line wings in the Balmer profiles (see Fig. 1 in Hessman et al. 1989) and the stronger absorption cores of the He I lines at $\lambda$4713 and $\lambda$4922 during outburst. These differences can be attributed to an increased optical thickness of the disc during outburst which would render the photospheric lines from the white dwarf undetectable (*i.e.* the broad absorption wings) while increasing the strength of the absorption core.

## 2.2 Light Curves

We rebinned the spectra on to a uniform wavelength scale and averaged them into 10 binary phase bins. We subtracted each individual continuum from each spectrum using a fitting procedure and the final result is displayed in Fig. 2. There is orbital variation in the line profiles, most notably in H$\beta$. The blue peak of the profile is clearly stronger than the red peak between phases 0.4-0.7. For these phases, the minimum of the absorption core has a redward offset of ~100 km s$^{-1}$ relative to the centre of the line. In addition, we observe that the emission peak closer to zero velocity is the stronger one (blue or red, depending on phase). During eclipse (−0.03 to +0.03 orbital phase; in quiescence both white dwarf and bright spot are eclipsed; Wood et al. 1989a), the width and height of the profile peaks are identical within the errors (using Gaussian profile fits). However, the gaussian centres of the peaks are −510±50 and +860±50 km s$^{-1}$ which indicate a redward $\gamma$ velocity shift for the eclipse profile. Such velocity shifts have been linked to sinusoidal components present in the trailed spectra (Hessman et al. 1989; see also next section of this paper).

The eclipse spectrum shows clear evidence of a double-peaked profile for He I $\lambda$4922 and possibly for He II. The H$\beta$ eclipse profile shows a symmetrical double-peaked profile with the absorption core above the continuum. Since the inner disc is eclipsed, the velocity separation of the two peaks can give an estimate of the outer disc radius, assuming Keplerian velocities, from the equation

$$\frac{R_{disc}}{a} = (1 + q) \left(\frac{K_r}{V_{min}}\right)^2$$

where $2 V_{min}$ = 1390±30 km s$^{-1}$ from Fig. 2, (typical of high inclination systems; Bailey & Ward 1981). For the above calculation we used results from Wood & Horne (1990; $K_r$ = 460 km s$^{-1}$, for a limb-darkening parameter of $0 < u < 1$ and $q = 0.102$).

The above gives $R_{disc}$ = 0.48±0.02 $a$, which is larger than the bright spot radius during quiescence, $R_{bs}$ = 0.31±0.02 $a$ (where $a$ is the binary separation; Wood et al. 1989a). This is consistent with a disc radius increase during outburst (Wood et al. 1989b) although sub-Keplerian velocities in the outer disc may also contribute to a difference between the bright spot radius and the emission line region (Wade & Horne 1988).

The equivalent width of H$\beta$ is around 7-8 Å and drops slightly to 5-6 Å at orbital phase 0.9. During eclipse it rises up to 75Å. Thus, the effect of the orbital hump is not very evident in the light curve of the H$\beta$ equivalent width and can account at most for 25 per cent. Although the emission flux calibration is not very accurate, no significant emission line flux in H$\beta$ is evident from the orbital hump.

### 2.3 The Trailed Spectra

We display the trailed spectra in Fig. 3 after having subtracted the continuum with a spline function. Close inspection of the double-peaked spectra shows various features caused by (a) the eclipse, (b) a sinusoidal wave from the red star and (c) an 'S'-wave.

During eclipse, the classical rotational disturbance or 'z'-wave is seen (blue peak is first eclipsed and then emerges first out of the eclipse), evidence of a prograde accretion disc been eclipsed (Greenstein & Kraft 1959). During eclipse, the absorption component disappears between phases 0.984-1.012 and is partially eclipsed (defined by the absorption core reaching the continuum level) at phases 0.957-0.984 and 1.012-1.039. We can constrain the size of the absorption region from the phases that it is observed to be fully eclipsed. The phases over which the white dwarf is fully eclipsed are from 0.980 to 1.020 (see Fig. 3 from Wood et al. 1989a) which are consistent with the range over which the absorption component disappears (i.e. comparable size), and thus its origin is related to the material which lies in the line of sight to the white dwarf.

A sharp, sinusoidal, emission component which moves in the opposite direction to the one of the double-peaked profile (i.e. the accretion disc) is also evident. In particular, it crosses the Balmer peaks (red to blue) at around phase 0.5, suggesting that its origin is related to the secondary star. More evidence on this is given in the following section.

It is possible to trace part of another sinusoidal-like, emission component in Fig. 3 with a larger velocity amplitude than the emission component we have already discussed.

It appears strong on the red peak after the eclipse (0.1-0.2 phase or ~36 degrees), then disappears and re-emerges strongly in the blue peak at phases 0.5-0.6, crosses the profile at phase ~0.75 and thereafter can be traced back to eclipse. It is apparent that these characteristics are signs of an 'S'-wave trailing the motion of the red star by about 0.25 cycles (see also next section). More evidence of the presence of the 'S'-wave is shown in the next section.

The 'S'-wave disappears when it crosses from red to blue suggesting that it is either eclipsed by an optically thick region or that it is seen against an optically thick background (so that an absorption component would counter balance the emission one). Considering the former, a vertically extended bulge (*i.e.* bright spot) and its aspect during phases 0.2-0.5 could in principle eclipse the 'S'-wave. However, imaging of the line distribution does not show any significant emission from a bright spot (see next section) which does not give support for any extended vertical structure.

## 3. Doppler Tomography

We used the trailed spectra of the H$\beta$ Balmer line to apply the indirect imaging technique of Doppler tomography (Marsh and Horne 1988). This is a method of reconstructing the emission-line distribution in a cataclysmic variable from the line-profile variation with orbital phase. In a Doppler image, the intensity of a pixel at a velocity ($V_X$, $V_Y$) corresponds to the intensity of the appropriate sinusoidal component in the trailed spectra

$$V = \gamma - V_X \cos(2\pi\phi) + V_Y \sin(2\pi\phi)$$

The semiamplitude of each sinusoidal component is given by the distance from the mass centre, while its phase is estimated from the azimuthal angle from the line-of-centres in the map. For example, emission from the red star follows $V_R = K_R \sin(2\pi\phi)$, and appears on the image at (0,$K_R$). Examples of the application to real data are given in Marsh & Horne (1990), Marsh et al. (1990), Dhillon et al. (1994) and Harlaftis et al. (1994).

It was possible to apply the technique only to the H$\beta$ line because of signal-to-noise limitations. However, the H$\beta$ profiles after subtraction of the continuum have many negative values because of the absorption core and for the technique to work only positive values are allowed. Therefore, we added a Gaussian of FWHM = 700 km s$^{-1}$ to the data so that all have positive values, and then after the reconstruction of the image we subtracted the equivalent gaussian image in order to produce the final result. This procedure has no effect on the goodness of the fit or entropy when the latter is determined over small scales (Marsh et al. 1990). We further minimized $\chi^2$ (=1.5) by adjusting the systemic velocity which causes blurring on the image (Marsh & Horne 1988). We obtained $\gamma$=+21 km s$^{-1}$.

Fig. 4 shows the results of this analysis for OY Car. The trailed spectra (54 out-of-eclipse spectra in the orbital range 0.1–0.9) of H$\beta$ are displayed in top panel "a" (with the velocity relative to the line centre along the horizontal axis and orbital phase along the vertical axis). Top panel "b" shows the Doppler map fitted to the observed data. From this map we subtract the axisymmetric Doppler map centred on the white dwarf (0, $-q\ K_R$). The result is presented in bottom panel "b" which shows the asymmetries in the line distribution more clearly. The bottom panel "a" shows the data corresponding to the latter image (non-axisymmetric emission); it is the result after a Gaussian profile (representing the axi-symmetric image) has been subtracted from the spectra. All panels of Fig. 4 have been plotted in the same intensity scale (0–0.9 mJy).

From top to bottom, the predicted positions of the red star, the centre of mass and the white dwarf are marked as crosses. The positions were determined for $K_R$=460 km s$^{-1}$ (0< $u$ <1), and $q = 0.102$ (Wood & Horne 1990). The two curves plotted at the bottom panel "b" and starting from the red star represent the velocity of the gas stream (lower curve) and the Keplerian velocity of the disc along the path of the gas stream. These curves are plotted for half orbit of the stream. The open circles denote distance from the white dwarf in units of 0.1 $R_{L_1}$. The asterisk marks the turning point of the velocities from high to low (closest to white dwarf) and the open circles next to the asterisk mark a distance of only 0.1 $R_{L_1}$ from the white dwarf. Top panel "c" shows the fits computed from the Doppler map (predicted data) and bottom panel "c" shows the fits computed from the non-axisymmetric Doppler map in the same grey scale as the observed data. The computed data (panel "c", overall and asymmetric parts) show that the observed data can be reconstructed reasonably well, including the red star component and the parts of the 'S'-wave where emission is strong (the red peak at phases 0.1-0.2 and blue peak at phases 0.5-0.6). The gap in the trailed spectra corresponds to the phases 0.62–0.67 when the flux standard star was observed (the eclipse was left out of the fit because it represents a geometric effect).

The double-peak profiles, characteristic of emission-line formation in an accretion disc, are evident in the trailed spectra. Reconstruction of the emission-line distribution shows this as a ring; the inner side of the ring (low velocities) corresponds to emission from the outer disc and the outer side of the ring (high velocities) to the region of the accretion disc near the compact object. Other than the dominant accretion disc emission, a narrow emission component is also evident in the trailed spectra. It crosses the double-peaked profile at phase 0.5 (from the red peak) in anti-phase with the orbital motion. The Doppler map determines the location of this emission-line component to be on the red star and in particular on its inner face.

In addition to the red star emission, there is extended, arm-like emission in the non-axisymmetric map (*i.e.* after the symmetric disc emission has been subtracted). It starts from the red star and actually spreads around the Roche lobe of the red star. This is not noise as we can reconstruct the same non-axisymmetric map by binning the spectra by a factor of 3 (thus decreasing the resolution to 120 km s$^{-1}$). However, it is unclear if it is related to the red star or the gas stream.

The main arm-like emission between the red star (1.0 $R_{L_1}$) and the 0.5 $R_{L_1}$ point lies close to the velocities of the gas stream or more accurately between the velocities of the gas stream and the disc tracking the stream. Moreover, close to 0.5 $R_{L_1}$ along the gas stream there are two spots (other than the red star) in the Doppler map which if related to the bright spot would give a disc size consistent with the size of the emission line region estimated from the double-peak separation in eclipse.

The arm-like emission after the 0.5 $R_{L_1}$ point does not follow the gas stream velocity. It follows a path similar, but one of lower velocities (by ∼200 km s$^{-1}$), to the path of the disc velocities along the trajectory of the stream. The emission extends along half the orbit of the gas stream around the white dwarf and the map indicates that the gas stream penetrates the disc as close as 0.1 $R_{L_1}$ from the white dwarf (the asterisk marks the turning point of the trajectory).

We estimate the contribution of the secondary star to the H$\beta$ line emission from the non-axisymmetric map to be between 2.4 per cent (counting only pixels covering the Roche lobe of red star) to a minimum of 0.3 per cent (counting only pixels covering the inner face of the red star; assuming that there is underlying emission from the gas stream we obtain a lower limit by subtracting it). The total non-axisymmetric emission accounts for ∼9 per cent of the total H$\beta$ emission which indicates that the gas stream emission is between 7–9 per cent.

A direct consequence of the detection of the secondary star from its Balmer emission during outburst is an estimate of its radial velocity which can thus provide an alternative estimate of the K$_r$ binary parameter. In this case where the geometry of the system is well determined we can check the consistency of the above proposed method. The peak line emission (between 360–480 km s$^{-1}$) has its average at $K_V = 420\pm80$ km s$^{-1}$ (each pixel is 40 km s$^{-1}$), consistent with the photometric value of 455< $K_r$ <470 km s$^{-1}$ (Wood & Horne 1990). It is clear that the limit of this spectroscopic method is the instrumental resolution (80 km s$^{-1}$ in our case). Note, also that the peak line emission $K_V = (0, V)$ is between the L$_1$ point and the centre of mass of the red star, on the side of the star that faces the white dwarf. Then, the use of $K_V$ for the determination of the true $K_r$ of the red star, would have caused an under-estimate in the sense that $K_r$ would be short of ∼35–50 km s$^{-1}$ of its real value.

## 4. Discussion

The H$\beta$ line emission is shown to be concentrated towards the inner hemisphere of the red dwarf in OY Car whereas this is not the case for the red star in IP Peg where the line emission is almost symmetrically distributed around the pole (Harlaftis et al. 1994, Marsh & Horne 1990). Geometrical considerations such as the larger size of the red star and the larger (*i.e.* thicker) disc in IP Peg in comparison to OY Car could in principle explain the above difference.

Balmer emission from the secondary star has been observed from about ten dwarf novae in outburst and nova-like variables. Recently, Balmer emission from the red star was detected during outburst in YZ Cnc as well. A Doppler tomography analysis of observations that covered 3 outbursts of YZ Cnc (Harlaftis et al. 1995) showed that a small portion of the H$\alpha$ emission originates on the secondary star, lasting throughout the outburst but dropping with the continuum decline. Marsh and Horne (1990) found in IP Peg Balmer emission from the gas stream during quiescence, and Balmer emission from the secondary star during outburst, in addition to the dominant disc emission. They argued that during outburst this emission originates from the chromosphere of the secondary which is irradiated by the boundary layer between the white dwarf and the inner accretion disc. Soft X-ray and EUV radiation from the boundary layer that is not absorbed by the accretion disc will reach the red star. This effectively means that only the polar regions of the red star will be irradiated, the extent of which will depend on the disc's thickness which can thus be investigated. The weak He II 4686 emission shows a double-peak profile (Fig. 2) which suggests that the red star may be irradiated by the disc. The line emission from the bright spot if not absent (see non-axisymmetric map) is very weak to account for any irradiation of the inner face of the red star.

The above suggests that either the disc is very thin during outburst so that there is no significant screening of the flux irradiating the red star or that the emission on the inner face of the red star is produced from a mechanism intrinsic to the star. The maximum line flux of H$\beta$ from the red star is 0.4 x $10^{30}$ erg s$^{-1}$ (the flux density of H$\beta$ is 16.54 mJy from Fig. 1). When compared with an accretion rate of $10^{-9}M_\odot yr^{-1}$ (Z Cha in outburst; Horne & Cook 1985) the luminosity from the boundary layer is roughly 1 x $10^{34}$ erg s$^{-1}$ with only 1% intercepted by the red star. Assuming an albedo of 0.5 for the red star, the H$\beta$ irradiated line flux is less than the total irradiating flux by about 2 orders of magnitude which shows that there is enough energy for irradiation to work.

The non-axisymmetric Doppler map indicates that the velocity of the gas stream changes when it finds the accretion disc (from velocity close to ballistic to a velocity closer to keplerian). The line distribution shows that the disc is penetrated from the gas stream to a distance very close to the white dwarf which is seen even after the turning point of the trajectory. The above suggests that the path of the gas stream is not significantly disturbed from its interaction with the accretion disc. This can be accommodated if the gas stream is thicker from the disc which would make it spill over the accretion disc.

The case of OY Car where emission from the gas stream and the red star is observed simultaneously during outburst is the first one and its implications on the state of the red star (intrinsic activity, a relation to the outburst mechanism and/or irradiation from a hotter source) causes an interest for observing a similar outburst in the near future. The origin of the line emission on the red star, which has been seen in systems of all types and outburst states (e.g. IP Peg in outburst by Marsh & Horne 1990 and in quiescence by Harlaftis et al. 1994), has yet to be resolved.

*Acknowledgements.* We thank Keith Horne and Frank Bateson for informing us about the outburst. ETH was in receipt of a HUMAN CAPITAL AND MOBILITY PROGRAM fellowship, funded by the European Union and administered by the Royal Observatories, UK. TRM is supported by a PPARC Advanced Fellowship.

**Fig. 1.** The average spectrum of OY Car out-of-eclipse showing the complex emission lines during outburst

**Fig. 2.** The spectra of OY Car in outburst are shown with orbital phase (after their continuum has been subtracted; this simplifies comparison with the eclipse spectrum). The data have been averaged into 10 bins and have been offset in the y-direction for clarity by 6 mJy each. The dominant lines are marked. Note the emission flux in H$\beta$ is maximum and the absorption component is minimum during eclipse

**Fig. 3.** The trailed spectra of H$\beta$ in OY Car (79 AAT spectra). The line He I 4921 is also displayed. The continuum has been subtracted from the spectra. Note the eclipse, the narrow sinusoidal component crossing red to blue at around phase 0.5 and parts of an 'S'-wave. See the text for more details

**Fig. 4.** The trailed spectra of H$\beta$ in OY Car are displayed in panel "a" (top). Panel "b" (top) shows the Doppler map fitted to these data. Panel "c" (top) shows the fits computed from the Doppler map (predicted data). Panel "b" (bottom) displays the non-axisymmetric component of emission. The bottom panels "a" and "c" show the observed and computed non-axisymmetric spectra, respectively. All panels are in the same grey scale. See text for more details